\documentclass{WileyMSP-template}
\usepackage{hyperref}
\usepackage[labelfont=bf]{caption}
\usepackage[superscript,biblabel]{cite}
\usepackage{setspace}
\begin{document}
\begin{sloppypar}

\pagestyle{fancy}
\rhead{\includegraphics[width=2.5cm]{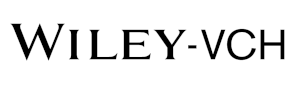}}

\title{ZrNb(CO) RF superconducting thin film with high critical temperature in the theoretical limit}

\maketitle


\author{Zeming Sun$^{\dagger,*}$}
\author{Thomas Oseroff$^{\dagger,*}$}
\author{Zhaslan Baraissov}
\author{Darrah K. Dare}
\author{Katrina Howard}
\author{Benjamin Francis}
\author{Ajinkya C. Hire}
\author{Nathan Sitaraman}
\author{Tomas A. Arias}
\author{Mark K. Transtrum}
\author{Richard Hennig}
\author{Michael O. Thompson}
\author{David A. Muller}
\author{Matthias U. Liepe*}

$^{\dagger}${contributed equally}



\begin{affiliations}
Dr. Z. Sun, Dr. T. Oseroff, K. Howard, Prof. M. U. Liepe \\
Cornell Laboratory for Accelerator-Based Sciences and Education, Cornell University, Ithaca, NY, USA\\
Email Addresses: zs253@cornell.edu (Z.S.), teo26@cornell.edu (T.O.), mul2@cornell.edu (M.U.L.)  

Z. Baraissov, Prof. D. A. Muller\\
School of Applied and Engineering Physics, Cornell University, Ithaca, NY, USA

Dr. D. K. Dare\\
Cornell Center for Materials Research, Cornell University, Ithaca, NY, USA

Dr. N. Sitaraman, Prof. T. A. Arias\\
Department of Physics, Cornell University, Ithaca, NY, USA

Dr. B. Francis and Prof. M. K. Transtrum\\
Department of Physics \& Astronomy, Brigham Young University, Provo, UT, USA

A. C. Hire and Prof. R. Hennig\\
Department of Materials Science and Engineering, University of Florida, Gainesville, FL, USA

Prof. M. O. Thompson\\
Department of Materials Science and Engineering, Cornell University, Ithaca, NY, USA

\end{affiliations}



\begin{abstract}
Superconducting radio-frequency (SRF) resonators are critical components for particle accelerator applications, such as free-electron lasers, and for emerging technologies in quantum computing. Developing advanced materials and their deposition processes to produce RF superconductors that yield n$\Omega$ surface resistances is a key metric for the wider adoption of SRF technology. Here we report ZrNb(CO) RF superconducting films with high critical temperatures ($T_\mathrm{c}$) achieved for the first time under ambient pressure. The attainment of a $T_\mathrm{c}$ near the theoretical limit for this material without applied pressure is promising for its use in practical applications. A range of $T_\mathrm{c}$, likely arising from Zr doping variation, may allow a tunable superconducting coherence length that lowers the sensitivity to material defects when an ultra-low surface resistance is required. Our ZrNb(CO) films are synthesized using a low-temperature (100\,--\,200\,$^{\circ}$C) electrochemical recipe combined with thermal annealing. The phase transformation as a function of annealing temperature and time is optimized by the evaporated Zr-Nb diffusion couples. Through phase control, we avoid hexagonal Zr phases that are equilibrium-stable but degrade $T_\mathrm{c}$. X-ray and electron diffraction combined with photoelectron spectroscopy reveal a system containing cubic $\beta$-ZrNb mixed with rocksalt NbC and low-dielectric-loss ZrO$_2$. We demonstrate proof-of-concept RF performance of ZrNb(CO) on an SRF sample test system. BCS resistance trends lower than reference Nb, while quench fields occur at approximately 35\,mT. Our results demonstrate the potential of ZrNb(CO) thin films for particle accelerator and other SRF applications.
\end{abstract}


\section{Introduction}

\begin{figure}[t!]
\centering
\includegraphics[width=\linewidth]{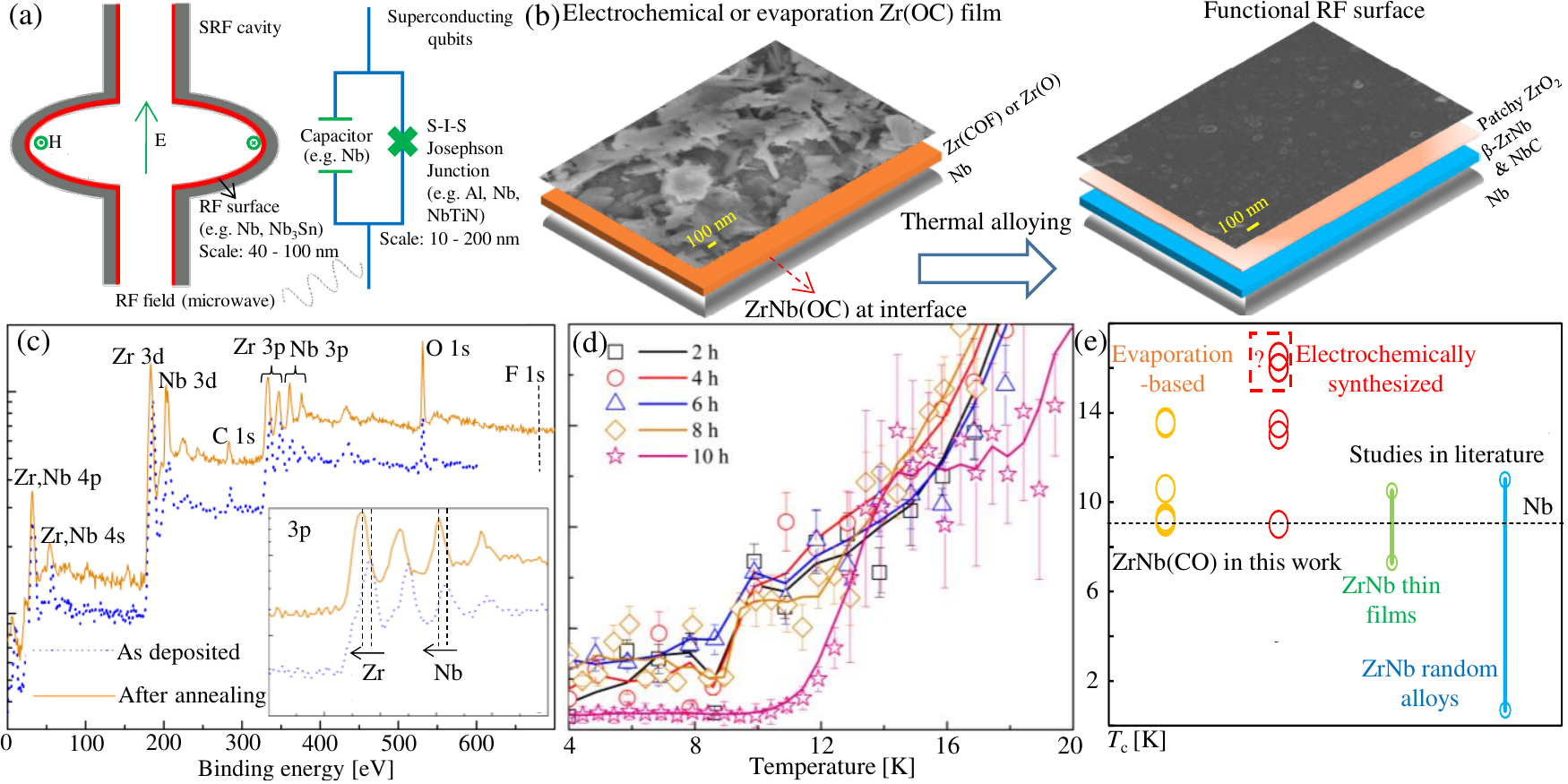}
\caption{\textbf{ZrNb(CO) alloying via electrochemical synthesis and thermal annealing to enable functional SRF surfaces.} (a) SRF cavity and quantum computing qubit, showing the use of SRF thin films. (b) Schematic showing the alloying process and surface morphology. (c) XPS spectra taken after removing surface oxides on samples as-deposited versus thermally annealed. (d) Resistivity drop measurements to determine $T_\mathrm{c}$ for electrochemical samples made via 2\,--10\,h synthesis followed by annealing, with a note on the potential impact of wire bonding on measurements for 2\,--\,8 h deposited nanometer-thin films. (e) $T_\mathrm{c}$ comparison between the electrochemical ZrNb(CO) and evaporated Zr-Nb diffusion couples with experimental results from literature\cite{SunRef32,SunRef33,SunRef35,SunRef52,SunRef53,SunRef54,SunRef55}.}
\label{SunFig1}
\end{figure}

Superconducting radio-frequency (SRF) resonators are critical components for particle accelerator applications, including synchrotrons and free-electron lasers (the fourth-generation light source)\cite{SunRef93,SunRef94,SunRef95,SunRef1,SunRef2}, high-energy particle colliders \cite{SunRef3}, nuclear physics experiments\cite{SunRef96}, dark-matter detection \cite{SunRef4}, materials/chemistry/biology research\cite{SunRef95,SunRef97,SunRef6}, and medical/biopharmaceutic applications \cite{SunRef6}, alongside technologies in $<$\,5\,nm semiconductor device fabrication \cite{SunRef7} and quantum computing \cite{SunRef5,SunRef75,SunRef76}. SRF resonant devices, \textit{e.g.}, SRF accelerating cavities and superconducting quantum qubits (\textbf{Figure~\ref{SunFig1}}a), require low-temperature RF superconductors owing to their excellent superconducting and material properties that facilitate technology transfer to practical applications\cite{SunRef78}. 

\bigbreak

RF superconductors such as Nb, Nb$_3$Sn, NbN, NbTiN, NbTi, NbSi, NbRe, V$_3$Si, etc., although studied decades ago, have been revitalized when tailored to emerging SRF applications \cite{SunRef92,SunRef17,SunRef18,SunRef19,SunRef20}. These low-temperature superconductors can achieve ultra-low RF resistances (n$\Omega$-scale) in the vortex-free state. The active surface layer of SRF cavities or quantum qubits with excited RF fields has a depth of tens of nanometers, \textit {e.g.}, 40\,nm for conventional Nb (dictated by field penetration depth). Here, we aim to engineer a thin ZrNb(CO) film through electrochemical synthesis and post-processing to obtain the desired phase and composition, yielding a combination of superconducting parameters suitable for RF applications, particularly SRF cavities. 

\bigbreak

Since the adoption of superconducting technology, RF cavities are reliant on bulk Nb, \textit {e.g.}, for Linac Coherent Light Source II. However, Nb cavities, with a superheating field of 200\,mT and critical temperature ($T_\mathrm{c}$) of 9.2\,K, are reaching the theoretical limits of $\sim$\,50\,MV/m accelerating gradients and (4\,--\,6)\,$\times$\,10$^{10}$ quality factors at 2\,K, 1.3\,GHz operations \cite{SunRef13}. A quantum leap in SRF R\&D, involving high beam energy, high accelerating gradient, compact size, and low cost, requires searching for new SRF materials, with three primary goals:
\begin{itemize}
      \item Increasing $T_\mathrm{c}$ to lower the cost and complexity of cryogenic systems while maintaining RF performance;
      \item Reducing surface resistance, $R_\mathrm{s}$, and proportionally boosting quality factors, $Q_\mathrm{0}$, which minimize the cost-driving power dissipation;
      \item Enhancing the accelerating gradient, $E_\mathrm{acc}$, which is limited by the intrinsic superheating field, $B_\mathrm{sh}$, of the superconductor.
\end{itemize}

\bigbreak

Currently, Nb$_3$Sn is a strong candidate owing to its high predicted $B_\mathrm{sh}$ of $\sim$\,400\,mT (corresponding to $\sim$\,100\,MV/m $E_\mathrm{acc}$) and high $T_\mathrm{c}$ of 18\,K, allowing cost-effective 4.2\,K accelerator operations \cite{SunRef92}. Nb$_3$Sn cavities have now been developed into full-scale 9-cell operations \cite{SunRef16}. Other materials of interest, including NbN \cite{SunRef17}, NbTiN \cite{SunRef17,SunRef18,SunRef19}, V$_3$Si\cite{SunRef20}, and MgB$_2$ \cite{SunRef17}, are in the early R\&D stage. To date, among these alternative SRF materials, only Nb$_3$Sn has reached the phase of the first technology transfer to accelerator products\cite{SunRef16}. Still, current Nb$_3$Sn cavities suffer from the typical quench fields of 13\,--\,18 MV/m $E_\mathrm{acc}$ (with the highest value of 24\,MV/m \cite{SunRef16}), considerably lower than the predicted ultimate limit of $\sim$\,100\,MV/m \cite{SunRef16}. While ongoing atomic \cite{SunRef23} and theoretical \cite{SunRef25} analyses combined with the growth optimization of Nb$_3$Sn films\cite{SunRef27} continue, further efforts are called for to seek alternative materials and surface structures.

\bigbreak

Our motivation for exploring ZrNb alloying stems from several fundamental aspects. Recently, theorists within the Center for Bright Beams have discovered the exciting potential of ZrNb alloys for SRF, predicting high $T_\mathrm{c}$'s up to 17.7\,K and high $B_\mathrm{sh}$'s up to 350\,mT \cite{SunRef28,SunRef29}. These large values promise similar benefits as Nb$_3$Sn. Moreover, the ability to tune the Zr concentration profile in ZrNb allows for manipulation of $T_\mathrm{c}$ and coherence length ($\xi$), which can reduce sensitivity to defects \cite{SunRef30}. In contrast, Nb$_3$Sn has a small $\xi$ of 3\,nm resulting in strong sensitivity to material disorders, which is likely one of the limiting factors in reaching its theoretical performance. This experimental work, stimulated by the theoretical predictions \cite{SunRef28,SunRef29}, intends to achieve Zr substitutional doping in the Nb bcc lattice to increase $T_\mathrm{c}$ compared to Nb while maintaining reasonably large $B_\mathrm{sh}$'s and $\xi$'s, minimizing defect sensitivity, and facilitating the realization of enhanced cavity performance. 

\bigbreak

ZrNb random alloys, usually made by atmospheric arc, induction, or levitation melting, have been studied since the 1960s \cite{SunRef32,SunRef33,SunRef35}. These Zr-rich ($>$\,80\,$\%$) alloys are mainly used for magnet construction \cite{SunRef32}, nuclear structure \cite{SunRef36,SunRef38,SunRef39,SunRef40,SunRef41,SunRef43}, and biomedical \cite{SunRef37} applications. As for superconducting applications, prior to this work, the highest $T_\mathrm{c}$ obtained was 11\,K \cite{SunRef32, SunRef33, SunRef35}, lower than the predicted 17.7\,K $T_\mathrm{c}$ value \cite{SunRef28} in an ordered, cubic ($\beta$) structure. 

\bigbreak

Phase transformations are the dominant factor preventing ZrNb $T_\mathrm{c}$ from reaching the predicted maximum. Theoretical analysis indicates that locking the bcc $\beta$ phase that is not an equilibrium-stable phase is the route to obtaining a high $T_\mathrm{c}$ of 17.7\,K in ZrNb \cite{SunRef28}. This prediction is supported by $T_\mathrm{c}$ values measured under extreme conditions, \textit{e.g.}, at 2.8\,$\times$\,10$^5$ atmospheric pressure \cite{SunRef46}. Additionally, simulations on $\beta$-Zr obtained at 30\,GPa pressure show a high $T_\mathrm{c}$ of 17.1\,K, whereas the complexity of Zr phase transformations in experiments results in a 10\,K $T_\mathrm{c}$ at the same pressure \cite{SunRef51}. Our goal is to produce a stable $\beta$-ZrNb phase at room temperature and atmospheric pressure, thus translating theoretical predictions into practical SRF applications.

\bigbreak

High-temperature ZrNb solid solutions exist in the bcc $\beta$ phase (phase diagram \cite{SunRef49}) while interdiffusion coefficients are high, \textit{e.g.}, 10$^5$\,nm$^2$/s for Nb into pure Zr and vice versa 10$^{-1}$\,nm$^2$/s at 1000\,\textdegree C \cite{SunRef50}. However, the equilibrium-stable phase at room temperature is the hcp $\alpha$-Zr phase with a low $T_\mathrm{c}$ of 7\,K \cite{SunRef32,SunRef51}. Moreover, the metastable hcp $\omega$ phase with a 4\,K $T_\mathrm{c}$ \cite{SunRef51} frequently appears through quenching from above 1000\,\textdegree C temperatures (diffusionless $\beta$ $\rightarrow$ $\omega$), although the majority $\beta$ phase is frozen, or through 350\,--\,850\,\textdegree C low-temperature aging, $\textit{i.e.}$, $\beta$ $\rightarrow$ ($\beta$ + $\omega$) $\rightarrow$ ($\beta$ + $\alpha$) \cite{SunRef44,SunRef45}. Additionally, martensite $\alpha^\prime$-Zr is observed through rapid quenching (\textit{e.g.}, water cooling) in Zr-rich alloys, but it evolves into $\beta$ and $\alpha$ upon post-annealing, \textit{e.g.}, at 600\,\textdegree C \cite{SunRef41,SunRef42,SunRef36,SunRef37,SunRef40}. The $\alpha^\prime$-Zr phase is treated as an athermal $\omega$ phase by some researchers \cite{SunRef45}, but their reported diffraction angles differ\cite{SunRef36,SunRef37,SunRef42}.

\bigbreak

Table~S1 summarizes the phase transformations, heat treatments, and composition ranges for ZrNb random alloys reported in the literature \cite{SunRef32,SunRef33,SunRef34,SunRef35,SunRef36,SunRef37,SunRef38,SunRef39,SunRef40,SunRef41,SunRef42,SunRef43}. Clearly, the $\alpha$, $\omega$, and $\alpha^\prime$ phases are difficult to avoid in random alloys. For the conditions observing $\beta$ phases, one should be cautious about the limitation of phase determination techniques, especially concerning the low $T_\mathrm{c}$'s obtained. Moreover, phase separation is typical in the form of $\alpha$-Zr and $\beta$-Nb precipitates at grain boundaries \cite{SunRef32,SunRef36,SunRef41,SunRef42}. A single $\beta$-ZrNb phase is possible, as indicated by the shift of diffraction peaks in few studies \cite{SunRef37}. Instead of these well-studied random alloys, we adopt an electrochemical surface alloying process that leverages the unique reaction between Zr precursors and the parent bcc-Nb surface to achieve a substitutional bcc system, as illustrated in Figure~\ref{SunFig1}b.    

\bigbreak

Thin-film ZrNb has been produced by co-sputtering, yielding a $T_\mathrm{c}$ of 9.1\,K \cite{SunRef52}. The thickness modulation of thin Zr and Nb layers can enhance their $T_\mathrm{c}$ values to 10.5\,K\cite{SunRef53,SunRef54,SunRef55}. This enhancement is induced by the proximity effect, which maintains a high $T_\mathrm{c}$ if the dimension of low-$T_\mathrm{c}$ or even normal-conducting phases is smaller than the coherence length. Furthermore, phase stability studies show that bcc Zr is greatly favored over hcp Zr, owing to the smaller interfacial energy, as the dimension of Zr phases is reduced and the Nb fraction is increased for films with over 50\,at.\% Nb \cite{SunRef56}. Also, the improvement of bcc-ZrNb stabilization is observed in simulations of Nb-rich bulk alloys \cite{SunRef57}. Hence, our synthesis of inhomogeneous compositions with $>$\,50\% Nb fractions provides flexibility in uniformity while facilitating bcc phase control and achieving suitably high $T_\mathrm{c}$'s. (Note that the superheating field is not affected by this inhomogeneous design\cite{SunRef29}.)

\begin{figure}[bt]
\centering
\includegraphics[width=\linewidth]{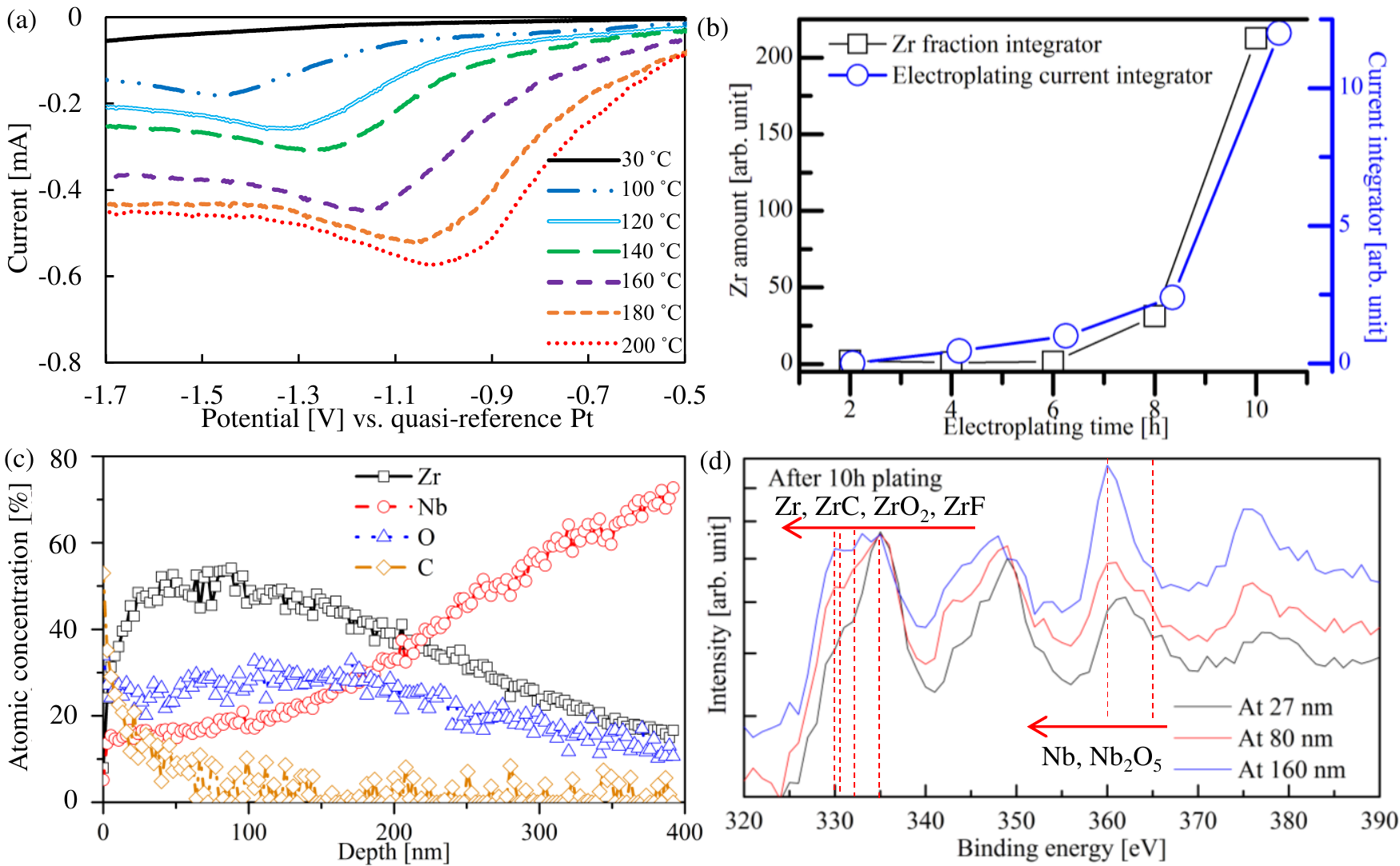}
\caption{\textbf{Electrochemical and physical analysis of Zr synthesis on Nb.} (a) Temperature-dependent CVs. (b) Total Zr amount and the corresponding plating current integrated over time, showing explosive Zr growth with onset at around 8\,h. (c,d) XPS elemental profiles (Zr, Nb, O, C) and 3p spectra (Zr, Nb) as a function of depth for electrochemical deposits prepared via 10\,h synthesis.}
\label{SunFig2}
\end{figure}

\bigbreak

Our electrochemical method meets the multi-scale needs ranging from deposition at the inner surface of a meter-long, complicated SRF cavity to the fabrication of nanoscale quantum devices. Previous investigations on Zr electrochemical deposition mainly focused on high-temperature (500\,--\,850\,\textdegree C) eutectic salts, as summarized in Table~S2 \cite{SunRef58,SunRef59,SunRef61,SunRef62,SunRef63,SunRef64,SunRef65,SunRef66,SunRef67,SunRef68,SunRef69,SunRef70}. However, these high temperatures are not practical for regular operations in an inert-gas glovebox, along with our concern of impurity doping induced \cite{SunRef72}. Indeed, low-temperature attempts are still missing. Moreover, chloride and chloride-fluoride studies suggest multiple reduction steps, unfortunately, prevail, \textit{e.g.}, Zr$^{4+}$$\rightarrow$Zr$^{2+}$$\rightarrow$Zr \cite{SunRef59,SunRef62,SunRef63,SunRef65,SunRef66,SunRef68}. Therefore, we have developed a low-temperature ($<$\,200\,\textdegree C) ionic-liquid process using fluorides that permit single-step reduction.       

\bigbreak

Here we present high-$T_\mathrm{c}$ ZrNb(CO) thin films produced using our electrochemical recipe, in combination with phase investigations using Zr-Nb diffusion couples. The inhomogeneous surface consisting of $\beta$-ZrNb and ZrO$_2$, mixed with rocksalt NbC in some samples, yields the highest-ever $T_\mathrm{c}$ values for this material system under ambient pressure. The highest values, as measured by temperature-dependent resistivity drops, are at least 13\,K (Figure~\ref{SunFig1}d and S16), with possible 16\,K values indicated by flux expulsion tests (Figure~S2 and uncertainty analyses in the Supporting Information). Our phase control and post-processing address the equilibrium constraints from hexagonal formations, leading to high $T_\mathrm{c}$ values that were previously only observed in theoretical simulations and extreme conditions.

\section{Results and Discussion}

\subsection{Zr Electrochemical Synthesis on Nb}

We have developed a low-temperature (100\,--\,200\,\textdegree C) Zr electroplating process on the Nb surface (See Experimental Section). We optimized the solution chemistry, reduction potentials, deposition temperature, and time using cyclic voltammetry (CV) and potentiostat. The CVs (Figure~S3) exhibit two reduction potentials at around -1 and -2.8\,V versus a Pt quasi-reference. By examining the deposits under scanning electron microscopy equipped with energy-dispersive X-ray spectroscopy (SEM/EDS) and X-ray photoelectron spectroscopy (XPS), we assigned the low reduction potential as Zr reduction and the high potential as electrolyte decomposition. The Zr reduction potential depends on the Zr$^{4+}$ concentration (following the Nernst equation) and the bath temperature that affects the Zr$^{4+}$ dissolution. As shown in \textbf{Figure~\ref{SunFig2}}a, the reduction shifts from -\,1.5\,V to -\,1\,V at increasing temperatures, along with an enlarged plating current. At increasing plating time under potentiostat, we observe an explosive Zr growth mode along with dendrite formation. Quantifying the total amount of Zr, probed by XPS and indicated by the integral of the plating current over the synthesis time (Figure~\ref{SunFig2}b), reveals rapid growth with an onset at around 8\,h. Surface SEM images (Figure~S4) show the nucleation of Zr sites at 2\,h and the formation of thin Zr films at 8\,h. The dendrites and a large amount of Zr deposition appear at 10\,h.

\bigbreak

Elemental analysis of the deposits, $\textit{e.g.}$, after 10\,h of synthesis (Figure~\ref{SunFig2}c and \ref{SunFig2}d), suggests that the alloying between Zr and Nb arises from reactive synthesis, generating the metallic Zr and Nb photoelectrons throughout the film, which matches the reactivity study\cite{SunRef79}. Oxygen appears within the deposits, and carbon accumulates only at the surface region. EDS spectra (Figure~S6a) show fluorine impurities that are later removed during thermal annealing (Figure~S6c); this behavior is confirmed by XPS (Figure~\ref{SunFig1}c). 

\bigbreak

The Zr and Nb 3p (Figure~\ref{SunFig2}d) and 3d (Figure~S7d) photoelectrons at different depths of the film show peak shiftings toward metallic motifs at increasing depths. Surface motifs, $\textit{e.g.}$, at a 27\,nm depth, may include zirconium carbides, oxides, fluorides, and niobium oxides in addition to metallic zirconium and niobium. Upon thermal annealing (Figure~\ref{SunFig1}c and S5), the Zr and Nb peaks further shift to the metallic motifs together with the formation of ZrO$_2$ and NbC. Since Zr typically shows strong gettering behaviors, we infer that impurity incorporation may not be fully avoided during synthesis, even under an inert-gas environment (O$_2$ and H$_2$O levels below 0.5\,ppm). Instead of further resolving impurity control, we focus on analyzing the impurity-induced phases in the final products after thermal annealing (see next section).

\subsection{Phase Transformation Mechanisms and High $T_\mathrm{c}$}

\begin{figure}[htbp]
\centering
\includegraphics[width=\linewidth]{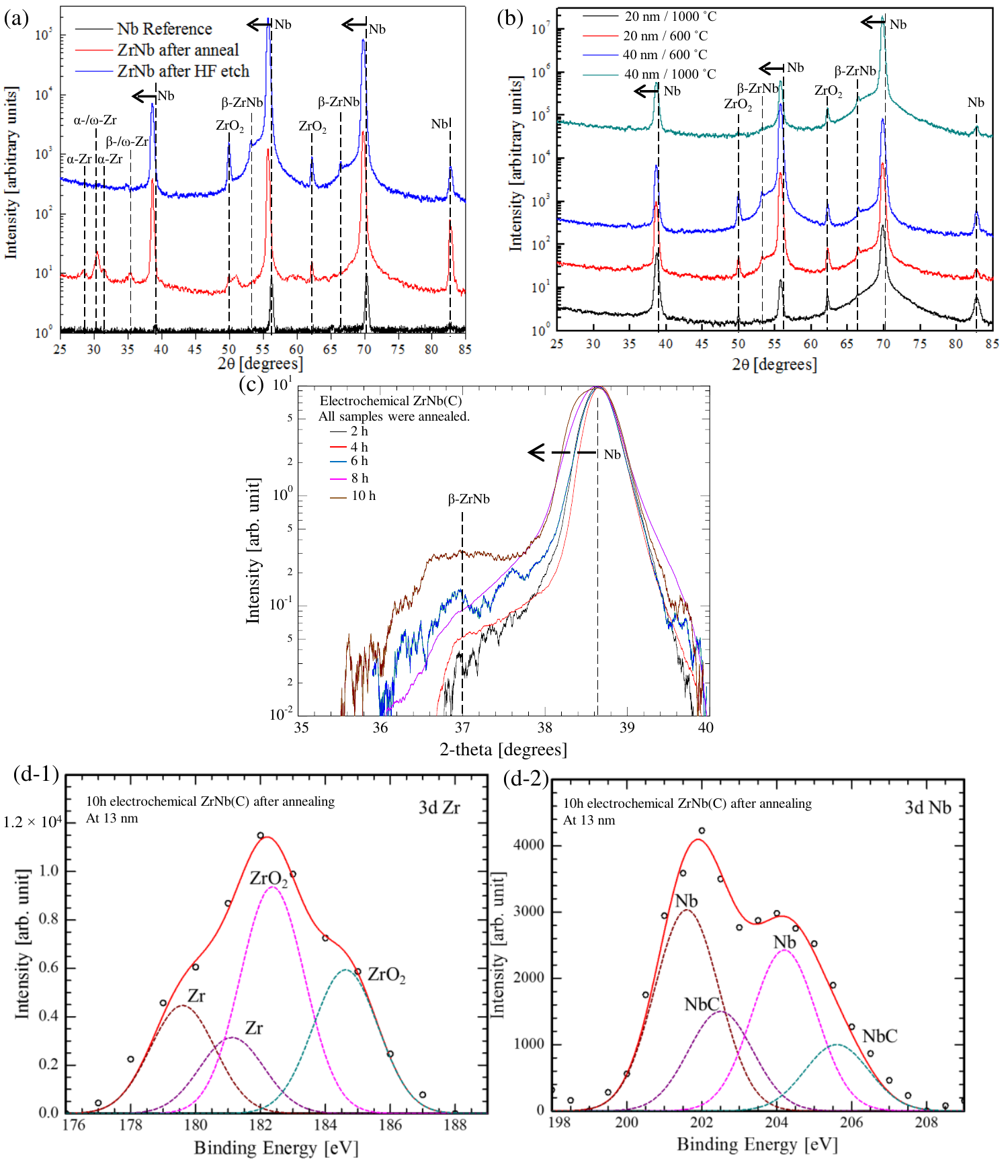}
\caption{\textbf{Phase analysis of Zr-Nb diffusion couples and electrochemically synthesized ZrNb(CO) after thermal annealing.} (a) Phase evolution of ZrNb diffusion couples after annealing at 600\,\textdegree C for 10\,h and after HF etch, compared to the Nb reference. (b) Phase comparison of ZrNb after annealing under different conditions for 10\,h. These samples underwent HF etching. (c) XRD patterns of the annealed ZrNb(CO) synthesized for varying times. (d) Deconvolution of XPS 3d spectra (Zr, Nb) for ZrNb(CO) prepared by 10\,h plating and annealed at 600\,\textdegree C for 10\,h. Note that the spectra shifted to lower binding energies by 0.6\,$\pm$\,0.1\,eV caused by the use of the neutralizer.}
\label{SunFig3}
\end{figure}

We first analyze the phase transformation after thermal annealing using evaporated Zr-Nb diffusion couples. Samples with initial Zr thicknesses of 20\,--\,40\,nm, which readily oxidized after evaporation (Figure~S14), were annealed under a vacuum of 2\,$\times$\,10$^{-7}$\,torr at temperatures of 600\,\textdegree C and 1000\,\textdegree C for durations of 20\,min to 10\,h. X-ray diffraction (XRD) patterns in Figure~S15 show the appearance of hcp $\alpha$- and $\omega$-Zr and ZrO$_2$, with indexing carefully referenced to the literature \cite{SunRef36,SunRef37,SunRef42,SunRef45,SunRef47}, as summarized in Table~S3. $\omega$-Zr is hard to distinguish as its diffraction angles are close to some $\alpha$- and $\beta$-Zr angles. After annealing, Zr doping peaks or shoulders appear on the lower 2$\theta$ regime next to the Nb peaks. Additionally, the Nb peaks slightly shift to lower 2$\theta$ values. Although we performed resistivity drop measurements using four probes on the surface of the annealed samples, we did not observe any $T_\mathrm{c}$ transition in the samples due to the presence of thick oxides on their surfaces.

\bigbreak

After etching the oxidized surface using HF acid, as shown in \textbf{Figure~\ref{SunFig3}}a, we notice the disappearance of hcp phases, regardless of $\alpha$-Zr or possible $\omega$-Zr. Moreover, the Zr doping peaks assigned to $\beta$-ZrNb become apparent, in addition to the Nb peak shifting. According to Bragg's and Vegard's laws, sin$\theta$\,$\propto$\,1/a in a cubic system, and a is linearly dependent on c, where $\theta$ is the diffraction angle, a is the lattice parameter, and c is the doping concentration. The substitutional Zr doping in a cubic Nb structure induces the diffractions between bcc Nb and Zr angles. Note that Zr's and Nb's bcc lattice parameters are 0.354\,nm and 0.330\,nm, respectively \cite{SunRef56}. These $\beta$-ZrNb diffractions are repeated for different annealing conditions with hexagonal Zr phases eliminated (Figure~\ref{SunFig3}b). As shown in Figure~S16, the diffusion couples made from 10\,h annealing followed by etching show $T_\mathrm{c}$ values ranging from 10.5\,--\,13.5\,K. In contrast, the 20\,min annealed samples show a niobium 9\,K $T_\mathrm{c}$ upon etching, indicating excessive etching on these samples. We find that the low-temperature anneals at 600\,\textdegree C yield higher $T_\mathrm{c}$ values than the 1000\,\textdegree C anneals. At 1000\,\textdegree C (see phase diagram\cite{SunRef49}), a thick region turns to a complete solid solution, and the subsequent cooling unalterably induces a large amount of $\alpha$-Zr precipitates, which degrade the $T_\mathrm{c}$. We, therefore, adopt the 600\,\textdegree C, 10\,h condition for annealing the electrochemical samples. 

\bigbreak

The annealed electrochemical ZrNb(CO) films show grain sizes of approximately 10\,--\,20\,nm (Figure~S8). XRD patterns of samples prepared from different synthesis times (Figure~\ref{SunFig3}c) exhibit the $\beta$-ZrNb phase with Nb peaks slightly shifting toward lower diffraction angles after annealing. XPS results further confirm the ZrNb alloying with distinctive metallic motifs at low binding energies (Figure~\ref{SunFig3}d). However, carbon and oxygen remain present (Figure~S7b and S7c), with oxygen existing throughout the film and carbon primarily appearing in the surface 20\,nm region. Under 4D-STEM, rocksalt niobium carbide is identified, as shown in Figure~S11. XPS also suggests the presence of NbC in the ZrNb film (Figure~\ref{SunFig3}d-2). Oxygen is fully bonded to Zr, as evidenced by the distinctive oxide peaks at high XPS binding energies (\textit{e.g.}, Figure~\ref{SunFig3}d-1). Nb oxides at 207\,--\,210\,eV are not observed (\textit{e.g.}, Figure~\ref{SunFig3}d-2), but Nb sub-oxides may exist. Pure ZrO$_2$ of 1\,--\,3\,nm exists on all samples, as shown in Figure~S10. This wide bandgap dielectric is a perfect choice for SRF applications. To sum up, we infer our electrochemical ZrNb(CO) includes $\beta$-ZrNb, NbC, and ZrO$_2$.

\bigbreak

The $T_\mathrm{c}$ results obtained from the 10\,h synthesis of electrochemical samples showed a value of 13\,K, as measured by the resistivity drop on $\sim$\,1\,cm$^2$ deposits (Figure~\ref{SunFig1}d). In contrast, during the flux expulsion test on 127\,cm$^2$ cavity-plate deposits, a small but consistent change in the magnetic field ($\sim$\,0.4\,mG) was observed at approximately 16\,K on three flux gates located at different positions, which \textit{suggest} that their $T_\mathrm{c}$ may be 16\,K; caution should be exercised in interpreting the flux expulsion results due to other possible explanations, as detailed in the Supporting Information (Figure~S2). 

\bigbreak

It is worth noting that both the sample-scale and cavity-plate depositions were performed using the same precursor concentration and reduction potential. However, the large-scale plate deposition had a 6 times smaller plating current density, likely due to the limited electrode size. To compensate for the reduced plating current, we extended the deposition duration to 29\,h. 
   
\bigbreak

The $T_\mathrm{c}$ values obtained in this work (10.5\,K\,--\,13\,K and possibly up to 16\,K) approach the theoretical limit of ZrNb \cite{SunRef28} and are close to those obtained under ultra-high pressures (28\,GPa) \cite{SunRef46}. The 13\,K $T_\mathrm{c}$ produced from the optimized condition is higher than the literature-reported values for ZrNb random alloys (11.5\,K\cite{SunRef32,SunRef33,SunRef35}) and sputtered ZrNb thin films (10.5\,K \cite{SunRef52,SunRef53,SunRef54,SunRef55}). The active phase determining $T_\mathrm{c}$ in the Zr-Nb diffusion couples (10.5\,--\,13.5\,K) should be $\beta$-ZrNb. In contrast, the active phase in the electrochemical ZrNb(CO), consisting of multi-phases, requires open discussion. 

\begin{figure}[t!]
\centering
\includegraphics[width=\linewidth]{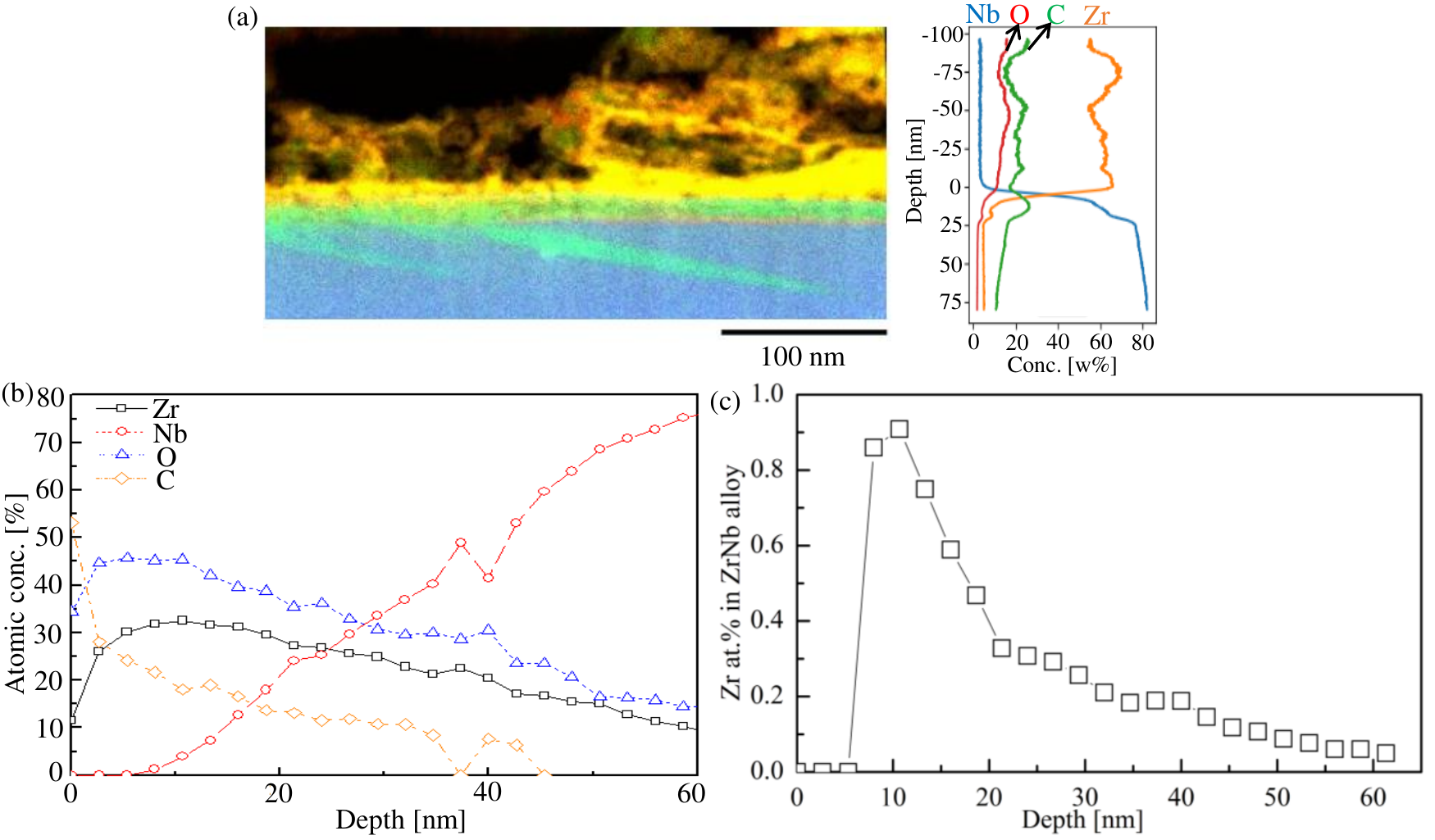}
\caption{\textbf{Composition profiles for the 10\,h electrochemical ZrNb(CO).} (a) Cross-sectional STEM-EDS elemental map and local depth profile. (b) XPS depth profiling. The probe size is $\sim$\,100\,$\mu$m. (c) Zr doping composition in $\beta$-ZrNb.}
\label{SunFig4}
\end{figure}

\subsection{Composition Analysis and Open Discussion on the Active Phase in ZrNb(CO)}

The cross-sectional EDS elemental mapping under STEM (\textbf{Figure~\ref{SunFig4}}a) and XPS depth profiling (Figure~\ref{SunFig4}b) reveal the composition of Zr, Nb, C, and O as a function of depth. The sputter-assisted XPS elemental compositions provide the relative quantities of these four elements spatially averaged over an approximately 100\,$\mu$m diameter probe. The local morphology with three-dimensional features (Figure~\ref{SunFig4}a) may affect this quantification, although XPS measurements at different sample locations show consistent depth profiles (Figure~S9a).

\bigbreak

By decomposing different motifs under the Zr and Nb XPS spectra ($\textit{e.g.}$, Figure~\ref{SunFig3}d), we quantified the relative fractions of metallic-Zr versus ZrO$_2$ and metallic-Nb versus NbC, as shown in Figure~S12. The surface layers are composed of ZrO$_2$. The film majority includes 30\% metallic Zr versus 70\% ZrO$_2$ and 65\% metallic Nb versus 35\% NbC. Note that the detection of ZrO$_2$ within the film may be from oxide precipitates or due to a residual signal induced by the surface roughness effect during ion sputtering. The nearly constant fractions over depth match the unaltered binding energy positions for Zr and Nb photoelectrons, despite their changing intensities at different depths (Figure~S9d). 

\bigbreak

The Zr doping concentration in $\beta$-ZrNb alloys is critical to their $T_\mathrm{c}$'s \cite{SunRef28} and superheating fields\cite{SunRef29}. We extract the Zr concentrations as a function of depth (Figure~\ref{SunFig4}c) in the $\beta$-ZrNb part of ZrNb(CO), after excluding the ZrO$_2$ and NbC signals. The profile reveals an inhomogeneous distribution, with a concentration spike with a small depth of $\sim$\,10\,nm, followed by a rapid decline reaching 10\,--30\,at.\% Zr with a depth of $\sim$\,20\,nm. Underneath these regions, the Zr concentrations remain $<$\,10\,at.\% and extend to a depth of $>$\,20 nm, consistent with the cross-sectional EDS depth profile (Figure~\ref{SunFig4}a). These low-concentration Zr values agree with the maximum solubility of $\sim$\,9\,at.\% Zr in the equilibrium $\beta$-ZrNb at 600\,\textdegree C. These results also match the $T_\mathrm{c}$ calculations \cite{SunRef28} predicting that the maximum $T_\mathrm{c}$ of 17.7\,K occurs at a Zr concentration of 25\,at.\%, and the $T_\mathrm{c}$ at 9\,at.\% Zr is 16\,K. 

\bigbreak

The question of whether the high $T_\mathrm{c}$ observed in this material system is determined by $\beta$-ZrNb or rocksalt NbC remains an open one. The literature reports that rocksalt NbC has $T_\mathrm{c}$'s of 9\,--\,11\,K for bulk phases \cite{SunRef80,SunRef81,SunRef82,SunRef83} and 8\,--\,11.5\,K for thin films \cite{SunRef84,SunRef85}, while rocksalt Nb$_x$Zr$_{1-x}$C, or Zr doping in NbC, reduces the $T_\mathrm{c}$ \cite{SunRef86}. However, DFT (density functional theory) simulations suggest that rocksalt and hexagonal NbC may have $T_\mathrm{c}$'s as high as 15.7\,K \cite{SunRef87} and 17.1\,K \cite{SunRef88}, respectively. Also, DFT simulations on the complex NbC$_{1-x}$N$_x$, or nitrogen doping in rocksalt NbC, predict $T_\mathrm{c}$'s ranging from 11.2\,K to 23.1\,K \cite{SunRef89}, suggesting that the effect of multiple light impurities on the $T_\mathrm{c}$ is not yet fully understood. 

\bigbreak

Our Zr-Nb diffusion couples containing $\beta$-ZrNb do not exhibit any detectable carbon signals, and these samples produce high $T_\mathrm{c}$'s close to those of ZrNb(CO). We argue that $\beta$-ZrNb is responsible for the observed $T_\mathrm{c}$ values above 11.5\,K, although we cannot rule out the possibility of high-$T_\mathrm{c}$ NbC.  

\subsection{Superconducting RF Performance}

\begin{figure}[!b]
\centering
\includegraphics[width=\linewidth]{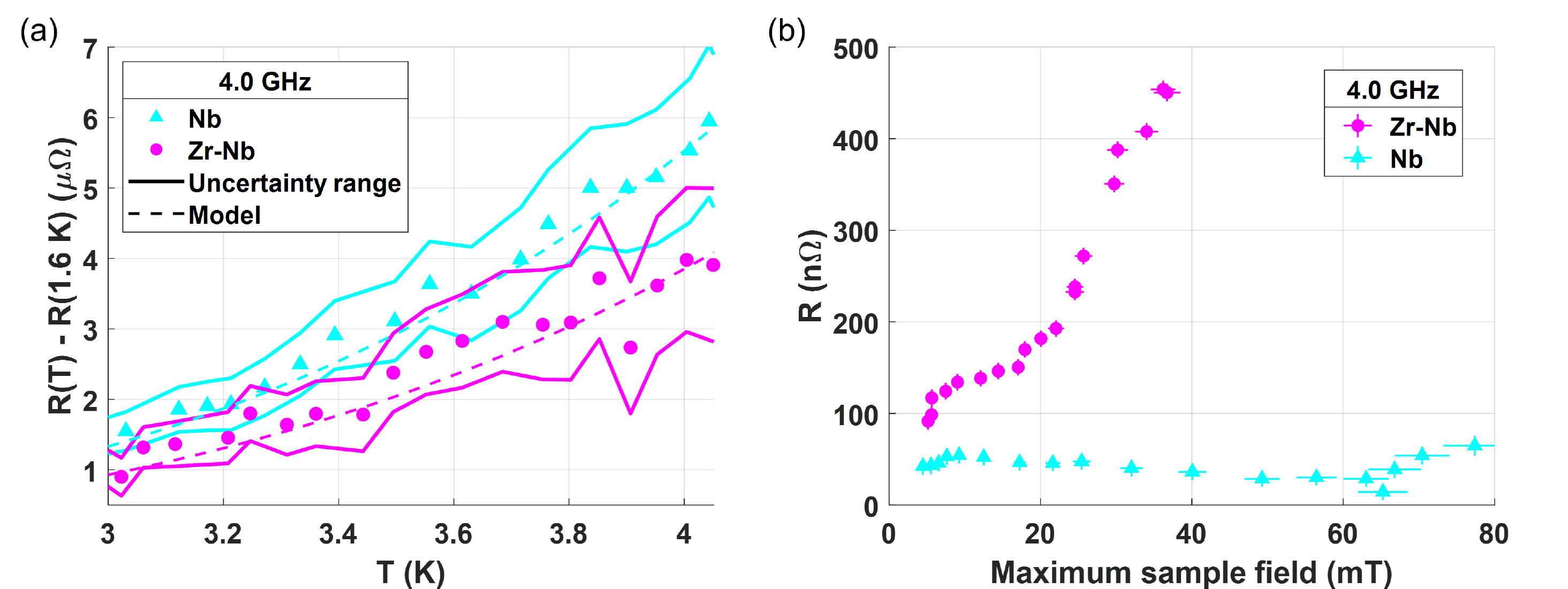}
\caption{\textbf{Superconducting RF performance of electrochemical ZrNb(CO).} (a) Measured surface resistances before (cyan triangles) and after (pink circles) the deposition of a ZrNb(CO) film onto a niobium substrate subtracted by its value at 1.6\,K as a function of temperature at 4.0\,GHz. R(1.6\,K, 4.0\,GHz) = 130\,n$\Omega$. The solid lines indicate the statistical uncertainty of the measurement. The cyan and pink colors indicate measurements of the niobium substrate before and after the growth of the ZrNb(CO) alloy. The dashed lines are model predictions from Equation (1) in Supplementary Information. (b) Measured total surface resistances at 1.6\,K and 4.0\,GHz comparing the field-dependence before (cyan triangles) and after (pink circles) the deposition of a ZrNb(CO) film onto a niobium substrate.  The highest field reported for each set corresponds to the maximum value before a quench event, above which the superconducting properties are lost.}
\label{SunFig5}
\end{figure}

We conducted a proof-of-concept RF demonstration of electrochemical ZrNb(CO) on the Cornell SRF sample test cavity \cite{SunRef73,SunRef90} (See Experimental Section). 

\bigbreak

The temperature-dependent surface resistances of the thin ZrNb(CO) alloy were measured at 4.0\,GHz and 5.2\,GHz with small RF field amplitudes of 2\,$\pm$\,0.6\,mT and 1\,$\pm$\,0.3\,mT, respectively, as the temperature was slowly decreased from 4.2\,K to 1.6\,K over several hours. To indicate the temperature-dependent BCS resistance, we subtracted from the measured surface resistances its value at 1.6\,K. \textbf{Figure~\ref{SunFig5}}a presents these surface resistances as a function of temperature at 4.0\,GHz and low amplitude fields (2\,$\pm$\,0.6\,mT). We found that the measured temperature-dependent BCS surface resistance trends towards lower values for ZrNb(CO) (pink circle) as compared to the reference Nb (cyan triangle), which is especially true at low fields and towards the higher end of the explored temperature range. This behavior was also observed at 5.2\,GHz. Using a simple model, we estimated the surface resistance of a bi-layer consisting of a 50\,nm film with a 16\,K $T_\mathrm{c}$ over bulk niobium (Supporting Information). The predicted results (dashed lines) aligned with the observed reduction in surface resistance. While this model and specified parameters were not intended to be a rigorous explanation of the results, this analysis provides a simple check that such behavior is consistent with a thin film having an elevated $T_\mathrm{c}$.  

\bigbreak

Low-temperature residual surface resistance with increasing field amplitude was measured for the ZrNb(CO) alloyed surface and the Nb reference (a baseline Nb that was later used for ZrNb(CO) deposition). The results at 1.6\,K and 4.0\,GHz are shown in Figure~\ref{SunFig5}b. The ZrNb(CO) residual surface resistance was found to rapidly increase with input power before reaching its quench field, \textit{i.e.}, the maximum RF magnetic field for which a flux-free Meissner state can exist. The quench fields were found to decrease from $\sim$\,80\,mT at 4.0\,GHz and $\sim$\,60\,mT at 5.2\,GHz with the original Nb substrate to $\sim$\,35\,mT for both frequencies after the addition of the ZrNb(CO) film. A curious feature of the ZrNb(CO) alloy surface resistance is an apparent independence on frequency, at least for the two studied here. This is an unusual behavior that is in contrast to the baseline measurement of the niobium substrate prior to the ZrNb(CO) growth.

\section{Conclusion}
In summary, we have observed promising characteristics of electrochemically deposited ZrNb(CO) thin films that are desirable for applications ranging from particle accelerators to quantum devices. The performance metrics include (i) generation of $\beta$-ZrNb, (ii) avoidance of low-$T_\mathrm{c}$ hcp Zr, (iii) high $T_\mathrm{c}$ (at least 13\,K) approaching the theoretical limit, and (iv) mild reduction of BCS resistance. This reduction could be more significant if the thickness of the ZrNb(CO) film were much larger than its penetration depth. This work marks the first step in developing electrochemical ZrNb(CO) and $\beta$-ZrNb for practical superconducting RF applications.


\section{Experimental Section}
\threesubsection{Electrochemical synthesis}\\

3\,mm thick Nb (RRR\,$>$\,300) substrates were electropolished via 9\,H$_2$SO$_4$\,:\,1\,HF with 100\,$\mu$m removal, and they were cleaned by 2\% HF before any deposition. The deposition was performed in an inert gas glovebox with O$_2$ and H$_2$O levels below 0.5\,ppm. ZrF$_4$ and LiF were dissolved in the ionic liquid 1-Butyl-1-methylpyrrolidinium bis(trifluoromethylsulfonyl)imide at different concentrations (0.48\,M\,--\,1.24\,M ZrF$_4$ and 0.98\,--\,4.8 times LiF addition). Clear solutions of light-yellow color were obtained upon $>$\,170\,\textdegree C heating, stirring, and standing, whereas the color turned brown during deposition. Chemicals were purchased from Krackeler Scientific, Inc., Alfa Aesar, and MilliporeSigma. A three-electrode electrochemical deposition system was employed, with Pt counter and quasi-reference wire electrodes. Electrochemical analysis was carried out using Princeton Applied Research VersaSTAT 3-500. Cyclic voltammetries were measured at scan rates of 10\,--\,50\,mV/s to identify reduction potentials, and the temperature was varied from 30\,--\,200\,\textdegree C using a hotplate with feedback control. A potentiostat controlled the deposition, and the deposition time ranged from 2\,--\,10\,h. After deposition, samples were cleaned with methanol, dried, and sealed in cleanroom plastic bags within the glovebox.

\threesubsection{Post annealing and processing}\\

The electrochemically synthesized samples were annealed in a Lindberg tube furnace at 600\,\textdegree for 10\,h under 2\,$\times$\,10$^{-7}$\,torr vacuum. 

\bigbreak

To identify the optimum annealing conditions, we initially investigated phase changes and diffusion properties using evaporated Zr-Nb diffusion couples. A Zr target was evaporated via e-beam on the electropolished bulk Nb surface at a 0.1\,$\AA$/s deposition rate under 1.3\,--\,1.8\,$\times$\,10$^{-6}$ torr base pressure. These films were readily oxidized (Figure~S14), while ZrO$_x$/ZrNb/Nb structures were formed after annealing, which matches the observation of ZrNb interfacial alloys for oxidized-Zr / Nb thin films in literature \cite{SunRef79}. Films of 20\,nm and 40\,nm thicknesses were annealed under $2\,\times\,10^{-7}$\,torr vacuum at 600\,\textdegree C and 1000\,\textdegree C for 20\,min\,--\,10\,h. We intended to design quenching and diffusion-dominated aging treatments through these annealing conditions. The heating rate was 5\,\textdegree C/min, and furnace cooling was used with average cooling rates of 10\,\textdegree C/min at 1000\,\textdegree C and 3\,\textdegree C/min at 600\,\textdegree C. After thermal annealing, samples were etched in 2\% HF for 30\,min to remove the surface ZrO$_2$ layer.

\threesubsection{Structural and elemental characterization} \\

A Zeiss Gemini scanning electron microscope (SEM) equipped with an in-lens detector under low voltage regimes (1\,--\,5\,kV) was used to evaluate surface morphology and uniformity. Cross-sections of the surface layers were imaged by FEI F20 and Thermo-Fisher Scientific Kraken scanning transmission electron microscopes (STEM) after specimen preparation using a Thermo Fisher Helios G4 UX Focused Ion Beam (FIB). Elemental information was obtained by X-ray photoelectron spectroscopy (SSX-100 XPS) depth profiling and by energy dispersive X-ray spectroscopy (EDS) under STEM and SEM. The ion sputter rate for depth profiling was measured using a standard gold film, and this value was further calibrated by comparing the sputter yields simulated from the SRIM software (Figure~S13) \cite{SunRef74}. Phase information was determined by a high-resolution Rigaku SmartLab X-ray diffractometer (XRD) and local electron diffraction under STEM. The XRD step size was 0.0012\textdegree, and the scan rate was 1.44\,\textdegree/min. Note that we attempted to minimize the samples' exposure to the atmosphere, \textit{e.g.}, using a vacuum puck or \textit{in situ} sample sealing, during transport from the glovebox to high or ultra-high vacuum techniques; still, short periods of exposure cannot be avoided.

\threesubsection{Superconducting property and RF performance} \\

The sample $T_\mathrm{c}$ was determined by temperature-dependent resistivity measurement conducted using a Quantum Design Physical Property Measurement System (PPMS) in the AC transport mode. Prior to the measurement, the sample surface was wedge-bonded to the sample puck using four 25\,$\mu$m aluminum wires (West Bond 747630E Wire Bonder).  

\bigbreak

The flux expulsion test was also used to indicate $T_\mathrm{c}$ by detecting changes in the magnetic fields surrounding the sample as it was warmed up from liquid-helium temperatures. When a superconductor transitions from the superconducting state to the normal conducting state, previously expelled magnetic flux re-enters the superconductor, resulting in an abrupt change in the magnetic fields in its vicinity, as depicted in Figure~S1. To conduct the tests, three flux gate magnetometers were attached to the back side of the Cornell sample test plate \cite{SunRef73,SunRef90} to monitor the magnetic fields while the sample was slowly warmed up at a rate of $\sim$\,0.3\,K/min. Each flux gate was oriented to measure the magnetic field along the axis normal to the sample surface, which is the dominant component of the fields in the cryostat, shielded to ambient values from $\sim$\,1\,mG to $\sim$\,10\,mG. The flux gates were positioned $\sim$\,2.5\,cm from the sample edge, arranged in the shape of a triangle. The tests were repeated twice, and further details are included in the Supporting Information.

\bigbreak

To evaluate the RF performance of ZrNb(CO), we scaled-up the electrochemical and annealing process to a 12.7\,cm diameter Nb plate that was fitted for the Cornell SRF sample test cavity \cite{SunRef73,SunRef90}. The cavity plate deposition used the same precursor concentration and reduction potential as the sample-scale deposition, but with a reduced plating current density due to the limited electrode size. To compensate, we extended the deposition time to 29\,h.

\bigbreak

We measured the surface resistances and quench fields of the plate with a thin surface layer of the ZrNb(CO) alloy at temperatures ranging from 1.6\,K to 4.2\,K using the sample test cavity. This modified cylindrical cavity operates with 4.0\,GHz TE$_{011}$-like and 5.2\,GHz TE$_{012}$-like modes. The sample plate was attached to the host structure as an end plate to the cylindrical opening, ensuring that the field on the sample was nearly identical for both modes of operation. The host structure geometry is designed to focus the surface fields on the location of the sample plate, allowing fields up to 100\,mT to be reached on the sample surface. The surface resistance of the sample plate was determined using a calibrated quality-factor measurement, with errors occurring when a sample has surface resistance comparable to that of the Nb host structure. For more information about this measurement setup and uncertainty analysis, see Reference [79].

\medskip
\textbf{Supporting Information} \par 
Supporting Information is available from the Wiley Online Library or from the author. The file includes a literature review on ZrNb phase transformation, diffraction, and Zr electrochemical deposition, as well as complete data from electrochemical synthesis and thermal annealing, Zr-Nb diffusion couples, and details on the estimation of surface resistance. 

\medskip
\textbf{Data availability} \par
All data generated or analyzed during this study are included in this published article and its supporting information file.

\medskip
\textbf{Author contributions} \par
Z.S. developed the methodology; conducted experiments on material growth, processing, optimization, characterization, and superconductivity measurements; and wrote the manuscript. T.O. conducted experiments and analyses on RF evaluation and flux expulsion test; and wrote the relevant section. Z.B. conducted experiments on cross-sectional STEM/EDS imaging and electron diffraction. D.K.D. assisted with XPS measurements, and K.H. assisted with part of the heat treatments. M.O.T., R.H., B.F., A.C.H., N.S., T.A.A., M.K.T., D.A.M., and M.U.L. provided valuable advice to design experiments and to understand results, mainly through the collaboration of the Center for Bright Beams. Z.S., T.O., Z.B., D.K.D., K.H., M.O.T., D.A.M., and M.U.L. further revised the manuscript. M.U.L. and D.A.M. acquired funding for the work and supervised this work.

\medskip
\textbf{Competing interests} \par
A provisional patent related to this research is filed by Cornell University. 

\medskip
\textbf{Acknowledgements} \par 
This work was supported by the U.S. National Science Foundation under Award PHY-1549132, the Center for Bright Beams. This work made use of the Cornell Center for Materials Research Shared Facilities which are supported through the NSF MRSEC program (DMR-1719875), and was performed in part at the Cornell NanoScale Facility, an NNCI member supported by NSF Grant NNCI-2025233. Z.S. thanks P. D. Bishop, A. Holic, J. Sears, G. Kulina, H. G. Conklin, and T. M. Gruber for helping with sample preparation and electrochemical system installation.
\medskip

%
\bibliographystyle{MSP}
\bibliography{Ref}

\begin{thebibliography}{10}
\providecommand{\url}[1]{\texttt{#1}}
\providecommand{\urlprefix}{URL }

\bibitem{SunRef93}
N.~Huang, et~al.,
\newblock \emph{The Innovation} \textbf{2021}, \emph{2} 100097, \protect{DOI}:
  \url{10.1016/j.xinn.2021.100097}.

\bibitem{SunRef94}
B.~W.~J. McNeil, N.~R. Thompson,
\newblock \emph{Nature Photonics} \textbf{2010}, \emph{4} 814, \protect{DOI}:
  \url{10.1038/nphoton.2010.239}.

\bibitem{SunRef95}
P.~G. O'Shea, H.~P. Freund,
\newblock \emph{Science} \textbf{2001}, \emph{292} 1853, \protect{DOI}:
  \url{10.1126/science.1055718}.

\bibitem{SunRef1}
W.~Decking, et~al.,
\newblock \emph{Nature Photonics} \textbf{2020}, \emph{14} 391, \protect{DOI}:
  \url{10.1038/s41566-020-00712-8}.

\bibitem{SunRef2}
E.~Prat, et~al.,
\newblock \emph{Nature Photonics} \textbf{2020}, \emph{14} 748, \protect{DOI}:
  \url{10.1038/s41566-020-0607-z}.

\bibitem{SunRef3}
V.~Shiltsev, F.~Zimmermann,
\newblock \emph{Reviews of Modern Physics} \textbf{2021}, \emph{93} 015006,
  \protect{DOI}: \url{10.1103/RevModPhys.93.015006}.

\bibitem{SunRef96}
C.~E. Reece,
\newblock \emph{Physics Review Accelerators and Beams} \textbf{2016}, \emph{19}
  124801, \protect{DOI}: \url{10.1103/PhysRevAccelBeams.19.124801}.

\bibitem{SunRef4}
R.~V. de~Water,
\newblock Presented at the 2022 North American Particle Accelerator Conference,
  Albuquerque, New Mexico, \textbf{2022} .

\bibitem{SunRef97}
W.~E. King, et~al.,
\newblock \emph{Journal of Applied Physics} \textbf{2005}, \emph{97} 111101,
  \protect{DOI}: \url{10.1063/1.1927699}.

\bibitem{SunRef6}
E.~J. Jaeschke, et~al.,
\newblock \emph{Synchrotron light sources and free-electron lasers: accelerator
  physics, instrumentation and science applications},
\newblock Springer, Cham, \textbf{2020}.

\bibitem{SunRef7}
W.~Ehrfeld, A.~Schmidt,
\newblock \emph{Journal of Vacuum Science \& Technology B: Microelectronics and
  Nanometer Structures Processing, Measurement, and Phenomena} \textbf{1998},
  \emph{16} 3526, \protect{DOI}: \url{10.1116/1.590490}.

\bibitem{SunRef5}
A.~Blais, et~al.,
\newblock \emph{Reviews of Modern Physics} \textbf{2021}, \emph{93} 025005,
  \protect{DOI}: \url{10.1103/RevModPhys.93.025005}.

\bibitem{SunRef75}
A.~Romanenko, et~al.,
\newblock \emph{Phys. Rev. Applied} \textbf{2020}, \emph{13} 034032,
  \protect{DOI}: \url{10.1103/PhysRevApplied.13.034032}.

\bibitem{SunRef76}
M.~H. Devoret, R.~J. Schoelkopf,
\newblock \emph{Science} \textbf{2013}, \emph{339} 1169, \protect{DOI}:
  \url{10.1126/science.1231930}.

\bibitem{SunRef78}
A.~Gurevich,
\newblock \emph{Nat. Mater.} \textbf{2011}, \emph{10} 255–259, \protect{DOI}:
  \url{10.1038/nmat2991}.

\bibitem{SunRef32}
J.~M. Corsan, et~al.,
\newblock \emph{Journal of Less-Common Metals} \textbf{1968}, \emph{15} 437,
  \protect{DOI}: \url{10.1016/0022-5088(68)90109-4}.

\bibitem{SunRef33}
J.~K. Hulm, R.~D. Blaugher,
\newblock \emph{Phys. Rev.} \textbf{1961}, \emph{123} 1569, \protect{DOI}:
  \url{10.1103/PhysRev.123.1569}.

\bibitem{SunRef35}
S.~Narasimhan, et~al.,
\newblock \emph{Journal of Nuclear Materials} \textbf{1972}, \emph{43} 258,
  \protect{DOI}: \url{10.1016/0022-3115(72)90057-8}.

\bibitem{SunRef52}
A.~Cavalleri, et~al.,
\newblock \emph{Journal of Physics: Condensed Matter} \textbf{1989}, \emph{1}
  6685, \protect{DOI}: \url{10.1088/0953-8984/1/37/015}.

\bibitem{SunRef53}
W.~P. Lowe, T.~H. Geballe,
\newblock \emph{Physical Review B} \textbf{1984}, \emph{29} 4961,
  \protect{DOI}: \url{10.1103/PhysRevB.29.4961}.

\bibitem{SunRef54}
P.~Koorevaar, et~al.,
\newblock \emph{Physical Review B} \textbf{1993}, \emph{47} 934, \protect{DOI}:
  \url{10.1103/PhysRevB.47.934}.

\bibitem{SunRef55}
U.~Gambardella, et~al.,
\newblock \emph{IEEE Transactions on Applied Superconductivity} \textbf{1993},
  \emph{3} 1253, \protect{DOI}: \url{10.1109/77.233397}.

\bibitem{SunRef92}
S.~Posen, et~al.,
\newblock \emph{Applied Physics Letters} \textbf{2015}, \emph{106}, 8 082601,
  \protect{DOI}: \url{10.1063/1.4913247}.

\bibitem{SunRef17}
T.~Oseroff, et~al.,
\newblock Proc. 19th Int. Conf. RF Supercond., Dresden, Germany, \textbf{2019}
  \protect{DOI}: \url{10.18429/JACoW-SRF2019-THP044}.

\bibitem{SunRef18}
Z.~Sun, et~al.,
\newblock Proc. 20th Int. Conf. RF Superconductivity, Lansing, MI, USA,
  \textbf{2021} \protect{DOI}: \url{10.18429/JACoW-SRF2021-WEPTEV012}.

\bibitem{SunRef19}
M.~C. Burton, et~al.,
\newblock \emph{Journal of Vacuum Science \& Technology A} \textbf{2016},
  \emph{34}, 2 021518, \protect{DOI}: \url{https://doi.org/10.1116/1.4941735}.

\bibitem{SunRef20}
K.~Howard, et~al.,
\newblock Proc. 20th Int. Conf. RF Superconductivity, Lansing, MI, USA,
  \textbf{2021} \protect{DOI}: \url{10.18429/JACoW-SRF2021-SUPFDV009}.

\bibitem{SunRef13}
A.~Grassellino, et~al.,
\newblock \emph{Superconductor Science and Technology} \textbf{2017},
  \emph{30}, 9, \protect{DOI}: \url{10.1088/1361-6668/aa7afe}.

\bibitem{SunRef16}
S.~Posen, et~al.,
\newblock \emph{Superconductor Science and Technology} \textbf{2021},
  \emph{34}, 2 025007, \protect{DOI}: \url{10.1088/1361-6668/abc7f7}.

\bibitem{SunRef23}
J.~Lee, et~al.,
\newblock \emph{Superconductor Science and Technology} \textbf{2018},
  \emph{32}, 2 024001, \protect{DOI}: \url{10.1088/1361-6668/aaf268}.

\bibitem{SunRef25}
A.~Pack, et~al.,
\newblock \emph{Phys. Rev. B} \textbf{2020}, \emph{101} 144504, \protect{DOI}:
  \url{10.1103/PhysRevB.101.144504}.

\bibitem{SunRef27}
Z.~Sun, et~al.,
\newblock Smooth, homogeneous, high-purity {Nb$_3$Sn RF} superconducting films
  by seed-free electrochemical synthesis, \textbf{2023},
\newblock \protect{DOI}: \url{10.48550/arXiv.2302.02054}.

\bibitem{SunRef28}
N.~Sitaraman, et~al.,
\newblock Theory of {Nb-Zr} alloy superconductivity and first experimental
  demonstration for superconducting radio-frequency cavity applications,
  \textbf{2022},
\newblock \protect{DOI}: \url{10.48550/arXiv.2208.10678}.

\bibitem{SunRef29}
B.~Francis, et~al.,
\newblock Superheating field of inhomogeneous surface layers in
  \protect{Ginzburg-Landau} theory using linear stability.

\bibitem{SunRef30}
M.~Tinkham,
\newblock \emph{Introduction to Superconductivity},
\newblock Dover Publications, \textbf{1996}.

\bibitem{SunRef36}
K.~Choo, et~al.,
\newblock \emph{Journal of Nuclear Materials} \textbf{1994}, \emph{209} 226,
  \protect{DOI}: \url{10.1016/0022-3115(94)90256-9}.

\bibitem{SunRef38}
G.~J. Cuello, et~al.,
\newblock \emph{Journal of Nuclear Materials} \textbf{1995}, \emph{218} 236,
  \protect{DOI}: \url{10.1016/0022-3115(94)00437-4}.

\bibitem{SunRef39}
G.~J. Cuello, et~al.,
\newblock \emph{Metall Mater Trans A} \textbf{2003}, \emph{35} 2771–2779,
  \protect{DOI}: \url{10.1007/s11661-003-0178-x}.

\bibitem{SunRef40}
S.~Banerjee, R.~Krishnan,
\newblock \emph{Acta Metallurgica} \textbf{1971}, \emph{19} 1317,
  \protect{DOI}: \url{10.1016/0001-6160(71)90068-X}.

\bibitem{SunRef41}
K.~Holm, et~al.,
\newblock \emph{Acta Metallurgica} \textbf{1977}, \emph{25} 1191,
  \protect{DOI}: \url{10.1016/0001-6160(77)90207-3}.

\bibitem{SunRef43}
H.~L. Yang, et~al.,
\newblock \emph{Journal of Nuclear Materials} \textbf{2016}, \emph{481} 117,
  \protect{DOI}: \url{10.1016/j.jnucmat.2016.09.016}.

\bibitem{SunRef37}
R.~Kondo, et~al.,
\newblock \emph{Acta Biomaterialia} \textbf{2011}, \emph{7} 4278–4284,
  \protect{DOI}: \url{10.1016/j.actbio.2011.07.020}.

\bibitem{SunRef46}
H.Kawamura, et~al.,
\newblock \emph{Physica B+C} \textbf{1984}, \emph{126} 485, \protect{DOI}:
  \url{10.1016/0378-4363(84)90217-1}.

\bibitem{SunRef51}
B.~T. Wang, et~al.,
\newblock \emph{Journal of Applied Physics} \textbf{2011}, \emph{109} 063514,
  \protect{DOI}: \url{10.1063/1.3556753}.

\bibitem{SunRef49}
H.~Okamoto,
\newblock \emph{Journal of Phase Equilibria} \textbf{1992}, \emph{13} 577,
  \protect{DOI}: \url{10.1007/BF02665776}.

\bibitem{SunRef50}
Y.~Liu, et~al.,
\newblock \emph{Calphad} \textbf{2008}, \emph{32} 455, \protect{DOI}:
  \url{10.1016/j.calphad.2008.06.008}.

\bibitem{SunRef44}
R.~F. Hehemann,
\newblock \emph{The Canadian Journal of Metallurgy and Materials Science}
  \textbf{1972}, 201--211, \protect{DOI}: \url{10.1179/cmq.1972.11.1.201}.

\bibitem{SunRef45}
G.~Aurelio, et~al.,
\newblock \emph{Journal of Nuclear Materials} \textbf{2005}, \emph{341} 1,
  \protect{DOI}: \url{10.1016/j.jnucmat.2004.12.001}.

\bibitem{SunRef42}
M.~Zhang, et~al.,
\newblock \emph{Journal of Alloys and Compounds} \textbf{2015}, \emph{651} 316,
  \protect{DOI}: \url{10.1016/j.jallcom.2015.08.105}.

\bibitem{SunRef34}
R.~F. Hehemann, S.~T. Zegler,
\newblock \emph{Trans. AIME} \textbf{1966}, \emph{236} 1594.

\bibitem{SunRef56}
G.~B. Thompson, et~al.,
\newblock \emph{Acta Materialia} \textbf{2002}, \emph{51} 5285, \protect{DOI}:
  \url{10.1016/S1359-6454(03)00380-X}.

\bibitem{SunRef57}
Y.~Zhao, et~al.,
\newblock \emph{Journal of Alloys and Compounds} \textbf{2021}, \emph{862}
  158029, \protect{DOI}: \url{10.1016/j.jallcom.2020.158029}.

\bibitem{SunRef58}
D.~Quaranta, et~al.,
\newblock \emph{Electrochimica Acta} \textbf{2018}, \emph{265} 586,
  \protect{DOI}: \url{10.1016/j.electacta.2018.01.213}.

\bibitem{SunRef59}
A.~Girginov, et~al.,
\newblock \emph{J. Appl. Electrochem.} \textbf{1995}, \emph{25} 993,
  \protect{DOI}: \url{10.1007/BF00241947}.

\bibitem{SunRef61}
H.~Groult, et~al.,
\newblock \emph{Journal of The Electrochemical Society} \textbf{2008},
  \emph{155} E19, \protect{DOI}: \url{10.1149/1.2811848}.

\bibitem{SunRef62}
L.~P. Polyakova, P.~T. Stangrit,
\newblock \emph{Electrochimica Acta} \textbf{1982}, \emph{27} 1641,
  \protect{DOI}: \url{10.1016/0013-4686(82)80092-3}.

\bibitem{SunRef63}
C.~Guang-Sen, et~al.,
\newblock \emph{J. Appl. Electrochem.} \textbf{1990}, \emph{20} 77,
  \protect{DOI}: \url{https://doi.org/10.1007/BF01012474}.

\bibitem{SunRef64}
B.-L. Yao, S.~Senderoff,
\newblock \emph{Journal of The Electrochemical Society} \textbf{2018},
  \emph{165} D6, \protect{DOI}: \url{10.1149/2.0211802jes}.

\bibitem{SunRef65}
F.~Basile, et~al.,
\newblock \emph{J. Appl. Electrochem.} \textbf{1981}, \emph{11} 645–651,
  \protect{DOI}: \url{10.1007/BF00616685}.

\bibitem{SunRef66}
R.~Baboian, et~al.,
\newblock \emph{Journal of The Electrochemical Society} \textbf{1965},
  \emph{112} 1221, \protect{DOI}: \url{10.1149/1.2423405}.

\bibitem{SunRef67}
G.~W. Mellors, S.~Senderoff,
\newblock \emph{Journal of The Electrochemical Society} \textbf{1966},
  \emph{113} 60, \protect{DOI}: \url{10.1149/1.2423865}.

\bibitem{SunRef68}
G.~J. Kipouros, S.~N. Flengas,
\newblock \emph{Journal of The Electrochemical Society} \textbf{1985},
  \emph{132} 1087, \protect{DOI}: \url{10.1149/1.2114020}.

\bibitem{SunRef69}
M.~A. Steinberg, et~al.,
\newblock \emph{Journal of The Electrochemical Society} \textbf{1954},
  \emph{101} 63, \protect{DOI}: \url{10.1149/1.2781210}.

\bibitem{SunRef70}
G.~M. Martinez, D.~E. Couch,
\newblock \emph{Metallurgical and Materials Transactions B} \textbf{1972},
  \emph{3} 575–578, \protect{DOI}: \url{10.1007/BF02642064}.

\bibitem{SunRef72}
F.~P. Lin, A.~Gurevich,
\newblock \emph{Phys. Rev. B} \textbf{2012}, \emph{85} 054513, \protect{DOI}:
  \url{10.1103/PhysRevB.85.054513}.

\bibitem{SunRef79}
Y.~S. Li, et~al.,
\newblock \emph{Applied Surface Science} \textbf{1996}, \emph{103} 389,
  \protect{DOI}: \url{10.1016/S0169-4332(96)00535-1}.

\bibitem{SunRef47}
A.~Jain, et~al.,
\newblock \emph{APL Materials} \textbf{2013}, \emph{1} 011002, \protect{DOI}:
  \url{http://link.aip.org/link/AMPADS/v1/i1/p011002/s1\&Agg=doi}.

\bibitem{SunRef80}
T.~Shang, et~al.,
\newblock \emph{Physical Review B} \textbf{2020}, \emph{101} 214518,
  \protect{DOI}: \url{10.1103/PhysRevB.101.214518}.

\bibitem{SunRef81}
D.~Yan, et~al.,
\newblock \emph{Physical Review B} \textbf{2020}, \emph{102} 205117,
  \protect{DOI}: \url{10.1103/PhysRevB.102.205117}.

\bibitem{SunRef82}
L.~E. Toth, et~al.,
\newblock \emph{Acta Metallurgica} \textbf{1965}, \emph{13} 379, \protect{DOI}:
  \url{10.1016/0001-6160(65)90064-7}.

\bibitem{SunRef83}
N.~K. Karn, et~al.,
\newblock \emph{Journal of Superconductivity and Novel Magnetism}
  \textbf{2021}, \emph{34} 2717–2724, \protect{DOI}:
  \url{10.1007/s10948-021-05994-9}.

\bibitem{SunRef84}
O.~MEYER, et~al.,
\newblock \emph{Thin Solid Films} \textbf{1973}, \emph{19} 217, \protect{DOI}:
  \url{10.1016/0040-6090(73)90057-6}.

\bibitem{SunRef85}
G.~Zou, et~al.,
\newblock \emph{Chem. Comm.} \textbf{2010}, \emph{46} 7837–7839,
  \protect{DOI}: \url{10.1039/c0cc01295e}.

\bibitem{SunRef86}
L.~E. Toth, et~al.,
\newblock \emph{Acta Metallurgica} \textbf{1966}, \emph{14} 1403,
  \protect{DOI}: \url{10.1016/0001-6160(66)90160-X}.

\bibitem{SunRef87}
E.~G. Maksimov, et~al.,
\newblock \emph{Journal of Experimental and Theoretical Physics Letters}
  \textbf{2004}, \emph{80} 548, \protect{DOI}: \url{10.1134/1.1846117}.

\bibitem{SunRef88}
X.~Li, et~al.,
\newblock \emph{Phys. Chem. Chem. Phys.} \textbf{2022}, \emph{24} 18419,
  \protect{DOI}: \url{10.1039/d2cp02403a}.

\bibitem{SunRef89}
S.~Blackburn, et~al.,
\newblock \emph{Physical Review B} \textbf{2011}, \emph{84} 104506,
  \protect{DOI}: \url{10.1103/PhysRevB.84.104506}.

\bibitem{SunRef73}
T.~Oseroff, et~al.,
\newblock Proc. 2021 International Conference on RF Superconductivity, Lansing,
  MI, USA, \textbf{2021} \protect{DOI}: \url{10.18429/JACoW-SRF2021-TUOFDV07}.

\bibitem{SunRef90}
T.~Oseroff,
\newblock Ph.D. thesis, Cornell University, \textbf{2022}.

\bibitem{SunRef74}
J.~F.Ziegler,
\newblock \emph{Nuclear Instruments and Methods in Physics Research Section B:
  Beam Interactions with Materials and Atoms} \textbf{2004}, \emph{219-220}
  1027.

\end{thebibliography}




\begin{figure} [h]
\textbf{Table of Contents}\\
\medskip
  \includegraphics{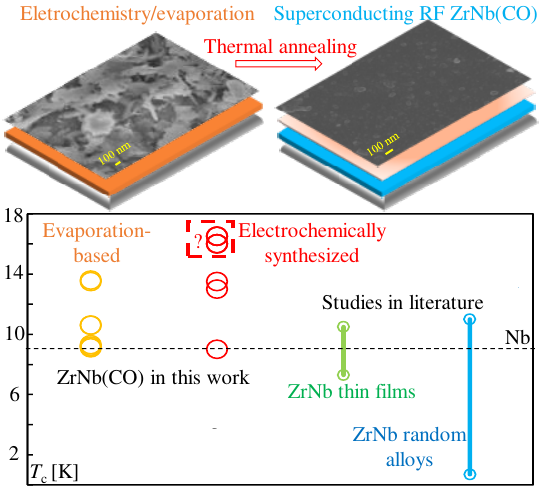}
  \medskip
  \caption*{Record-breaking critical temperatures are achieved in superconducting ZrNb(CO) films containing $\beta$-ZrNb without applied pressure. Controlled phase and composition via electrochemical synthesis and heat treatments produce desirable superconducting RF properties. This advancement facilitates the production of energy-efficient and compact SRF cavities for particle accelerators, ultra-sensitive electronics, and quantum applications.}
\end{figure}

\end{sloppypar}
\end{document}